\documentclass[apj]{emulateapj}
\usepackage{epsfig}
\usepackage{lscape}

\shorttitle{2MASS Galaxy Distribution Function}
\shortauthors{SIVAKOFF \& SASLAW}
\slugcomment{ }

\begin{document}
\title{The Galaxy Distribution Function from the 2MASS Survey}

\author{Gregory R. Sivakoff}
\affil{Department of Astronomy, University of Virginia,
P. O. Box 3818, Charlottesville, VA 22903-0818; grs8g@virginia.edu}
\and
\author{
William C. Saslaw
}
\affil{Department of Astronomy, University of Virginia,
P. O. Box 3818, Charlottesville, VA 22903-0818; Institute of Astronomy, Cambridge, England; and National Radio Astronomy Observatory\altaffilmark{1}, Charlottesville, VA; wcs@virginia.edu
}
\altaffiltext{1}{Operated by Associated Universities, Inc., under cooperative 
agreement with the National Science Foundation.}%
\setcounter{footnote}{1}

\begin{abstract}
We determine the spatial distribution function of galaxies
from a wide range of samples in the 2MASS survey. The results agree very
well with the form of the distribution predicted by the theory of
cosmological gravitational many-body galaxy clustering. On large
scales we find a value of the clustering parameter $b = 0.867 \pm 0.026$, in
agreement with $b = 0.83 \pm 0.05$ found previously for the Pisces-Perseus
supercluster. We measure $b(\theta)$ as a function of scale, since this is
a powerful test of the applicability of computer simulations. The results
suggest that when galaxies clustered they were usually surrounded by
individual, rather than by communal haloes.
\end{abstract}


\section{Introduction}
\label{sec:intro_2MASS_dist}

The spatial locations of galaxies may be described by many different
statistics. These include distribution functions, low order
correlation functions, multi-fractal dimensions, topological genus,
power spectra, spherical harmonics, multipoles, minimal spanning
trees, percolation and moments. All these statistics are related and
each emphasizes different aspects of the distribution (e.g.,
\citealt{S_2000}). Here we determine the observed galaxy distribution
function from the 2MASS Catalogs with considerably greater precision
than has previously been possible from smaller catalogs, and compare
it with theoretical expectations.

The galaxy distribution function contains information about
correlations to all orders, about voids and underdense regions, and
about clustering over a very wide range of intensities, sizes and
shapes. Moreover it can be calculated analytically for cosmological
gravitational many-body clustering in both linear and non-linear
regimes \citep{S_2000}.

These theoretical calculations assume, consistent with direct N-body
simulations, that galaxy clustering evolves through a sequence of
quasi-equilibrium states.
This occurs because in any overdense region
the local dynamical timescale is shorter than the global timescale for
evolution of average quantities, such as the overall density, velocity
dispersion, and the ratio of gravitational potential correlation
energy to the kinetic energy of peculiar velocities. The global
timescale for these average ensemble properties to change, becomes
even greater than the Hubble time, $R(t)/\dot{R}$, as local partially
virialized structures form. The Hubble expansion exactly cancels the
long-range gravitational many-body mean field, not only in the usual
Einstein-Friedmann models, but also in those models with a
cosmological constant and quintessence. For the cosmological constant
and quintessence models, one can straightforwardly extend the earlier
derivations \citep{SF_1996,S_2000} that applied to the Einstein-Friedmann
case.

Under these conditions, the clustering of galaxies, each of which may
be surrounded by its own dark matter halo, occurs mainly through their
mutual gravitational interactions. This differs from those cold dark
matter models that are dominated by very large haloes containing
galaxies moving mainly as test particles within the gravitational
field of these haloes (e.g, the massive haloes in \citealt{FM_2003}).
When the mutual gravitational interactions of
individual galaxies dominate, clustering can be described by
quasi-equilibrium thermodynamics \citep{SH_1984,SF_1996,S_2000} and
statistical mechanics (\citealt{ASB_2002}; \citealt{LS_2004}).
This theory applies for a wide range of initial conditions, including
initial power law perturbation spectra with $-1 \lesssim n \lesssim 1$
\citep{I_1990}.
Characterising the complete range of initial
conditions that form a basin of attraction for this theory remains an
important unsolved problem.

Analysis of the cosmological many-body problem predicted
\citep{SH_1984} that its distribution function $f_{V}(N)$, i.e. the
probability that a randomly placed volume $V$ in space contains $N$
galaxies, has the form
\begin{equation}
  \label{eq:GQED}
  f_{V}(N) = {{\overline{N}(1-b)}\over{N!}}[{\overline{N}(1-b)} +N b]^{N-1}
                                            e^{-[{\overline{N}(1-b)} +N b]},
\end{equation}
where
\begin{equation}
  \label{eq:N}
  \overline{N} = \overline{n} \ V
\end{equation}
is the expected number for an average number density $\overline{n}$.
The strength of clustering is measured by
\begin{equation}
  \label{eq:b}
  b = -\frac{W}{2K} = \frac{2 \pi G m^{2} \overline{n}}{3T}
                      \int_{V} \xi(\overline{n},T,r) \frac{1}{r} r^{2} dr
\end{equation}
for a spherical volume
where $W$ is the gravitational correlation energy, $K$ is the kinetic
energy of galaxy peculiar velocities indicating an ensemble average
temperature $T$, the average galaxy mass is $m$, and $\xi$ is the
two-galaxy correlation function. Since the volume $V$ may be a cone
with apex at the observer which projects the galaxies in its volume
into a cell of area $A$ on the sky, the same form of the distribution
function in equation~(\ref{eq:GQED}) applies to counts of galaxies in
two-dimensional cells on the sky. This form of equation~(\ref{eq:b})
assumes that evolution is negligible within the volume, which is reasonable for
a narrow redshift band as in the 2MASS catalog ($0 \le z \lesssim 0.1$). For
volumes covering wide redshift bands, evolution may be included as described
in \citet{FS_1997} and \citet{SE_2000}.

Although the form of equation~(\ref{eq:GQED}) is independent of the
shape and size of randomly located volumes which are not pathological
(e.g., not fractal or chosen to avoid galaxies), the value of $b$ will
depend on volume and shape. This was found in early N-body simulations
\citep{IIS_1988} and explicitly calculated for square projected cones
using equation~(\ref{eq:b}) \citep{LS_1992}. Comparisons of different
values of $\overline{N}$ and $b$ for cells of the same shape and size,
but for different subsamples of galaxies, can be used to explore
possible biases in these subsamples. If the bias is known {\it a
priori}, it can be incorporated into equations (\ref{eq:GQED}) and
(\ref{eq:b}) as described in detail previously
\citep{LS_1992,S_2000}. In principle, an unknown bias contained in
subsamples can also be recovered from the results, but its uniqueness
is more difficult to establish.

The value of $\overline{N}$ can be determined directly from the
catalog and the value of $b$ follows for equation~(\ref{eq:GQED}) from
the observed variance of counts in cells of any particular volume and
shape:
\begin{equation}
  \label{eq:var}
  \langle(\Delta N)^2_V \rangle = {{\overline{N}}\over{(1-b(V))^2}}.
\end{equation}
Therefore in principle the theory contains no free parameters for
comparing equation~(\ref{eq:GQED}) with the observations. We can also
fit $\overline{N}$ and $b$ in equation~(\ref{eq:GQED}) directly to the
observed distribution function, and see if the slight difference
between these two methods of determining $\overline{N}$ and $b$
provides useful information.

After deriving equation~(\ref{eq:GQED}), its predicted form was found
to agree well with counts in cells and void probabilities in several
galaxy surveys including the Zwicky Catalog \citep{SC_1991}, the UGC
and ESO Catalogs \citep{LS_1992} and the IRAS Catalog
\citep{SMS_1994}, all projected onto the sky, as well as for the SSRS
Catalog \citep{FZ_1994}, and the Pisces-Perseus Supercluster
\citep{SH_1998} in three dimensions. All previous comparisons have
involved catalogs containing at most several tens of thousands of
galaxies. The new catalogs becoming available have about 10--100 times
as many galaxies, with different levels of sky coverage, homogeneity,
and information about galaxy properties.

These new larger catalogs are important because their improved
statistics may reveal departures from equation~(\ref{eq:GQED})
resulting from different distributions of galaxies and dark matter,
merging, mass segregation, or other environmental effects.
Section~\ref{sec:analysis_2MASS_dist} describes the 2MASS Catalog and
our method of analysis, Section~\ref{sec:results_2MASS_dist} gives our
results, and Section~\ref{sec:discussion_2MASS_dist} describes some of
their implications.

\section{The 2MASS Catalog and its Analysis}
\label{sec:analysis_2MASS_dist}

The Two Micron All Sky Survey employed telescopes in the Northern and
Southern hemispheres to observe in the three infrared bands
J $(1.11 - 1.36 \mu m)$, H $(1.50 - 1.80\mu m)$ and
$K_{s} (2.00 - 2.32\mu m)$.
Most of the sources are points, but 1,647,599 are resolved with
respect to the observed 2MASS point spread function and they
constitute the Extended Source Catalog (XSC). About 97\% of these
extended sources are galaxies%
\footnote{The Explanatory Supplement to the 2MASS All Sky Data Release is
available at http://www.ipac.caltech.edu/2mass/releases/allsky/doc for this
and related information.\label{ftn:explanatory}}%
. Examination of the Abell 262 cluster suggests that most of
the 2MASS galaxies are local with $z \lesssim 0.1$ (Skrutskie, private
communication). 
In addition to the local population, the Catalog
contains more distant very luminous galaxies and about 0.4\% of the
catalog may be quasars with redshifts $\le 5$ \citep{BH_2001}. The
low redshifts for nearly all the sources indicate that evolutionary
corrections will not be significant for our statistics using this
sample.

To select our samples, we adopted Kron magnitudes following
\citet{MMK+_2005} and dereddened them using the reddening maps of
\citet{SFD_1998} assuming $A(J) = 0.82 E(B-V)$, $A(H) = 0.48 E_(B-V)$,
and $A(K_{s}) = 0.28 E(B-V)$ \citep{MSW+_2003}%
\footnote{The corrected magnitudes are denoted by a subscripted $0$.}%
. We required all galaxies brighter than $K_{s,0} = 12$ to have
$J_{0} - K_{s,0}$ between 0.7 and 1.4 \citep{MMK+_2005} and all
candidate galaxies to have star-galaxy separation indices typical of
galaxies ($e_{score} < 1.4$ and $g_{score}<1.4$) to avoid confusion with
nearby galactic sources.
We also require that a candidate is not
listed\footnotemark[\ref{ftn:explanatory}] as a possible anomaly
(e.g. artifact, cluster/galaxy piece, Galactic object or star) and
that if a candidate is among the $\sim 25\%$ visually verified
sources, it is not indicated as non-extended ($vc \neq 2$).
To minimize the relative contribution of stars in the samples, we
usually remove cells at low Galactic latitude, $|\delta_{gal}| < 20 \degr$.
In some cases we choose higher Galactic latitude cutoffs for comparison,
as shown in
Table~\ref{tbl:GQED} and Figure~\ref{fig:GQED_mag}. We also remove
cells with high stellar density, or in very dusty regions, as
described below. Although this does not explicitly avoid the
Magellanic clouds, the high stellar density cutoffs remove their
central regions.

To examine the completeness of the galaxy sample, we have measured
$\langle V/V_{max} \rangle$ as a function of $K_{s,0}$ magnitude. The results
are essentially constant between about 0.52 and 0.51 for $10.0 < K_{s,0} <
13.7$ and we employ a conservative truncation of the galaxy sample at
$K_{s,0} = 13.5$.

\begin{figure*}
\epsfig{file=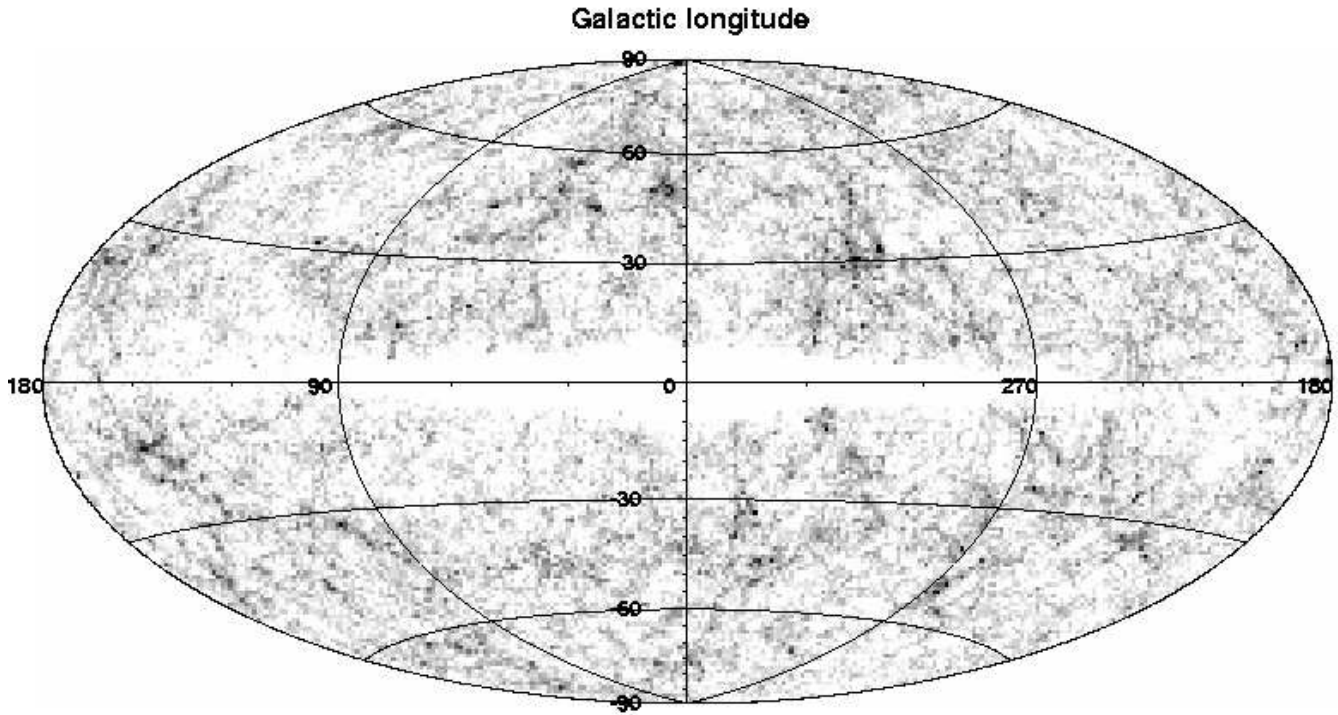,width=\textwidth,clip=}
\caption{
Grey-scale image of the density of 663,166 galaxy candidates with
$K_{s,0} < 13.5$ in cells of $1\degr \times 1\degr $. A square-root
stretch having a contrast of 2 was applied with the darkest cells indicating
the highest density.
\label{fig:GQED_allsky}}
\end{figure*}

To obtain the distribution function from counts-in-cells, we map the
sources onto a Hammer-Aitoff equal area projection (e.g., \citealt{CG_2002}):
\begin{eqnarray}
\gamma & = & \frac{180 \degr}{\pi}
             \sqrt{\frac{2}{1+\cos \delta \cos(\alpha/2)}}, \\
 x     & = & 2 \gamma \cos \delta \sin(\alpha/2), \nonumber\\
 y     & = & \gamma \sin \delta. \nonumber
\end{eqnarray}
Here ($\alpha$ , $\delta$) is the Galactic coordinate in radians and ($x$, $y$)
is the projected coordinate.
Figure~\ref{fig:GQED_allsky} shows the sample of galaxy candidates
with $K_{s,0} < 13.5$ on a square-root stretched grey scale with the
darkest cells indicating the highest density. The extended dense
region in the upper right hand (northwest) quadrant is the Shapley
Supercluster. This map contains 663,166 galaxy candidates in cells of
$1\degr \times 1\degr $on the sky.
We then apply two filters. The first is a spatial filter that removes
cells below a specified Galactic latitude or on the boundary of the
projection.
The second filter reduces the probability of confusion with stars by
removing cells whose stellar density, $n_{st}$, is greater than
$10^{3.3} (\approx 2000$~deg$^{-2})$ and cells with objects that have
$A(K_{s}) > 0.05$ \citep{MMK+_2005}.
We experimented with a range of stellar density
cutoffs and found $n_{st} < 10^{3.3}$ to be the optimum between leaving
too few galaxies or including so many stars that a small percentage of
them were close enough to each other that they could masquerade as
galaxies.

Having determined all the relevant cells, we simply count the number
of galaxies in each to construct histograms of the distribution
functions for various samples.
These filtered samples include
essentially the whole sky, four hemispherical sections, and four
quadrant sections, along with a range of cell sizes, Galactic latitude
cutoffs and magnitude limits. Values of $\overline{N}$ from the average
projected density and $b_{v}$ from the variance of counts in cells in
equation~(\ref{eq:var}) are used as a first approximation to determine
$f_{V}(N)$ in equation~(\ref{eq:GQED}) from the observations.
Then we go a step further and find values of $\overline{N}$ and $b$
that minimize a $\chi^{2}$ fit to the observations. The differences
in these values for a single histogram, and among similar histograms
in different areas of the sky, give a measure of their global
uncertainty. However values of $\chi^{2}$ cannot be used here for the
usual probability estimates that a sample represents a particular
distribution function because the populations of nearby cells are
often strongly correlated. Since these cells are not independent, the
values of $\chi^{2}$ are used just to optimize the fitting. To
facilitate this, we have combined bins with very small $f_{V}(N)$ in the
tails of the distribution (cf. \citealt{KS_1979}).

\section{Results}
\label{sec:results_2MASS_dist}

\begin{figure*}
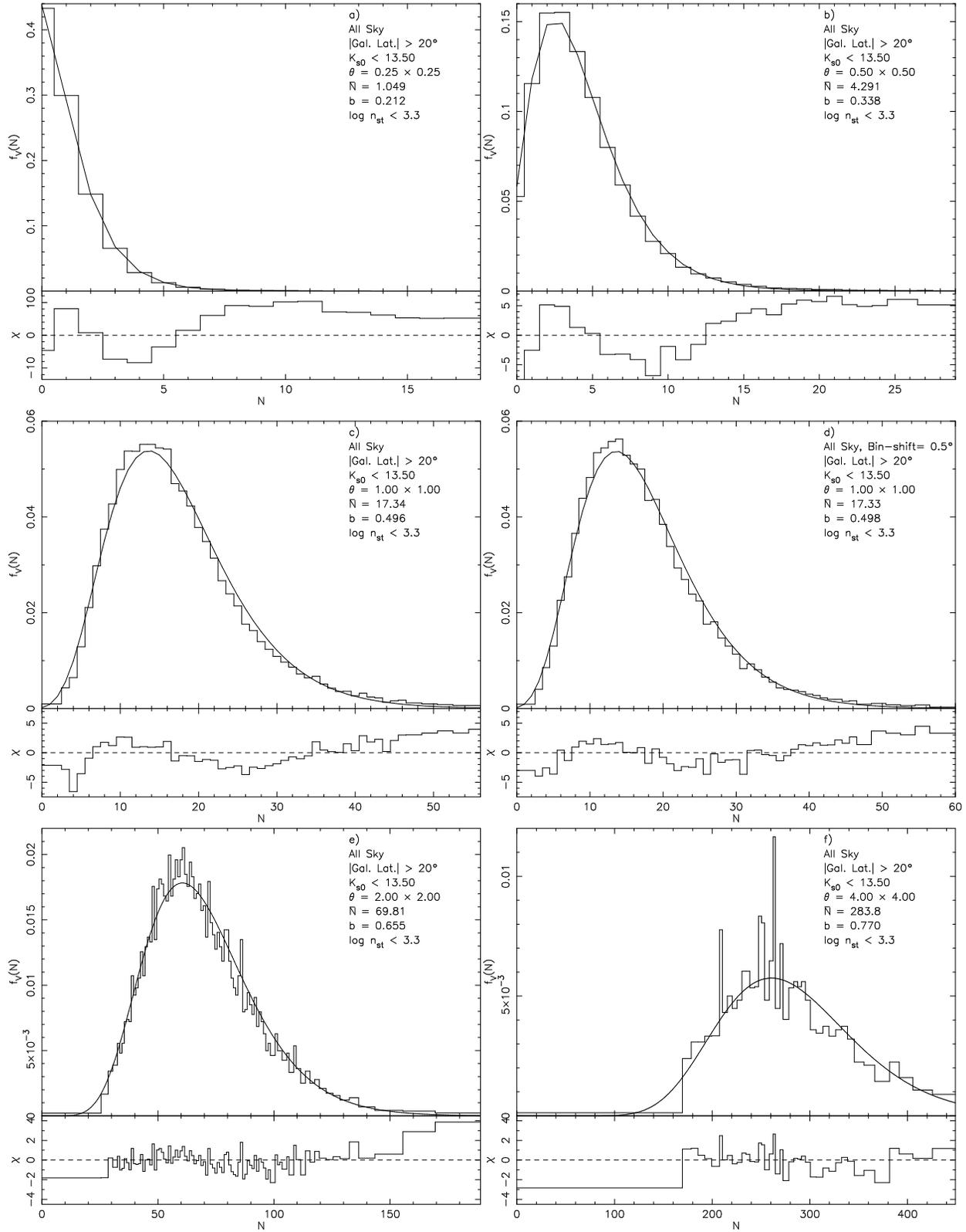

\epsfig{file=f2a.ps,angle=-90,clip=}\epsfig{file=f2b.ps,angle=-90,clip=}\\
\epsfig{file=f2c.ps,angle=-90,clip=}\epsfig{file=f2d.ps,angle=-90,clip=}\\
\epsfig{file=f2e.ps,angle=-90,clip=}\epsfig{file=f2f.ps,angle=-90,clip=}
\caption{
Least-squares fits (continuous lines) of $f_{V}(N)$ to the observed
histograms for square cells of six different size cells on the
sky. The area under the histograms in the bottom panel of each figure
displays the values of $\chi$ for the fit. The plots exclude the final
(high $N$'s) bin. In (d), the histogram was measured after shifting
the bins by $0.5\degr$ in longitude.
\label{fig:GQED_size}}
\end{figure*}

To illustrate the relation between the predicted form of $f_{V}(N)$ in
equation~(\ref{eq:GQED}) and the observed 2MASS galaxy distribution
function, Figures~\ref{fig:GQED_size}a-~\ref{fig:GQED_size}f plot
least-squares fits of $f_{V}(N)$ to the observed histograms for square
cells of five different size cells on the sky.
The areas under these histograms in the bottom panel of each figure illustrate
values of $\chi = (f_{obs} - f_{theory}) N_{cell}^{1/2} / f_{obs}^{1/2}$,
where $N_{cell}$ is the total number of cells evaluated.
The $1\degr \times 1\degr$ cells in Figures~\ref{fig:GQED_size}c and
\ref{fig:GQED_size}d are representative; in Figure~\ref{fig:GQED_size}d,
we have shifted the grid of cells by $0.5\degr$ on the sky. This
shift slightly alters the detailed shape of the histogram for the
probability that a cell has $N$ galaxies, but leaves the best fit
values of $\overline{N}$ and $b$ almost unchanged.
Thus the exact positioning of the grid is unimportant, which indicates
that the galaxy distribution is statistically homogeneous.

In the overall fits there are small differences between the
observations and the theoretical distribution of
equation~(\ref{eq:GQED}). The largest differences, as measured by
$\chi$, occur in the large $N$ tails where there are relatively few
cells. The peaks of the observed distribution function are usually
higher than those of equation~(\ref{eq:GQED}). Since the areas of the
distributions have the same normalization, a slight excess for some
values of $N$ must be offset by a deficit for other values. This
slight deficiency generally occurs between the peak and high $N$ tail
of the distribution. There may be several contributions to these
differences: the peculiarity of the unique configuration of large
clusters in the Universe, our emphasis on fitting the overall
distribution rather than giving greater weight to either the peak or
tail, very small systematic effects in the 2MASS catalog and its
analysis, or a real physical effect. If the last possibility is true,
it might be caused by several effects including environmental
interactions which tend to enhance the luminosities of very overdense
regions, mergers in relatively overdense regions which enhance the
number of cells around the peak of the distribution, a remnant of
initial conditions when galaxies started clustering, or segregation of
more massive galaxies into smaller volumes. Since the overall
differences are so small, it will be a considerable challenge to determine
their causes.

\begin{figure}
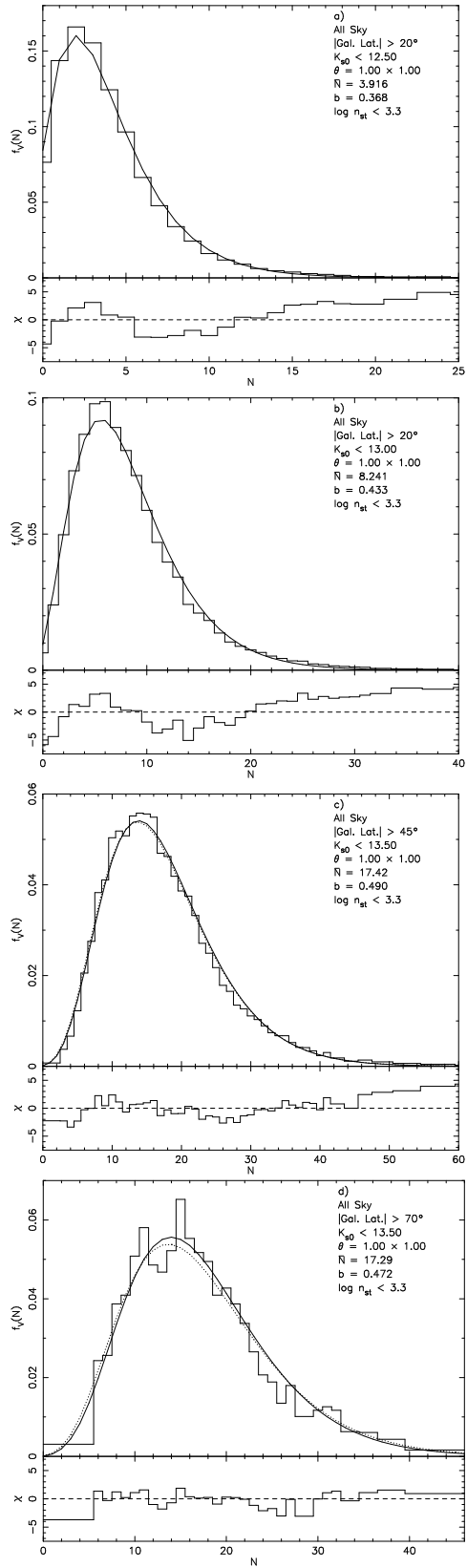

\epsfig{file=f3a.ps,angle=-90,clip=,width=2.50in}\\
\epsfig{file=f3b.ps,angle=-90,clip=,width=2.50in}\\
\epsfig{file=f3c.ps,angle=-90,clip=,width=2.50in}\\
\epsfig{file=f3d.ps,angle=-90,clip=,width=2.50in}
\caption{
As Figure~\ref{fig:GQED_size}, except (a) and (b) indicate different
magnitude cuts while (c) and (d) indicate different
Galactic latitude cuts, all with $1\degr$ cells on the sky.
The dotted lines in (c) and (d) indicate the best-fit model from the
full sample with values of $b$ and $\overline{N}$ given in
Table~\ref{tbl:GQED}.
\label{fig:GQED_mag}}
\end{figure}

Figures~\ref{fig:GQED_mag}a-d show variations of these counts in
cells. The upper two comparisons are for $1\degr \times 1\degr$ cells
(which we will now use for succeeding representative illustrations)
over the entire sky, as before, but with magnitude cutoffs of $K_{s,0}
< 12.5$ and $K_{s,0} < 13.0$. Examining the values of $b$ for
different magnitude cutoffs can describe whether magnitudes are
partially correlated with position, as discussed in
Section~\ref{sec:discussion_2MASS_dist}. Comparison with
Figure~\ref{fig:GQED_size} shows that the excellence of the fit is not
much affected by the magnitude cutoff, as we would expect for a
statistically complete sample.
Figures~\ref{fig:GQED_mag}c and
\ref{fig:GQED_mag}d show samples with the standard magnitude cutoffs
but confined to Galactic latitudes more than $45\degr$ and $70\degr$
from the Galactic plane.
Comparison with the $1\degr \times 1\degr$ cells of
Figures~\ref{fig:GQED_size}c and
\ref{fig:GQED_size}d, shows that the fits are as good and the values
of $b$ are the same within 2\% for the $45\degr$ Galactic latitude cutoff
and within 4\% for the $70\degr$ cutoff. The $70\degr$ cutoff has larger
fluctuations because it has only 20\% of the number of cells as the
$45\degr$ cutoff and less than 10\% of the cells in the $20\degr$
cutoff. Evidently galactic obscuration is not a significant problem
above $\pm 20\degr$ latitude for these distribution functions.

\begin{figure}
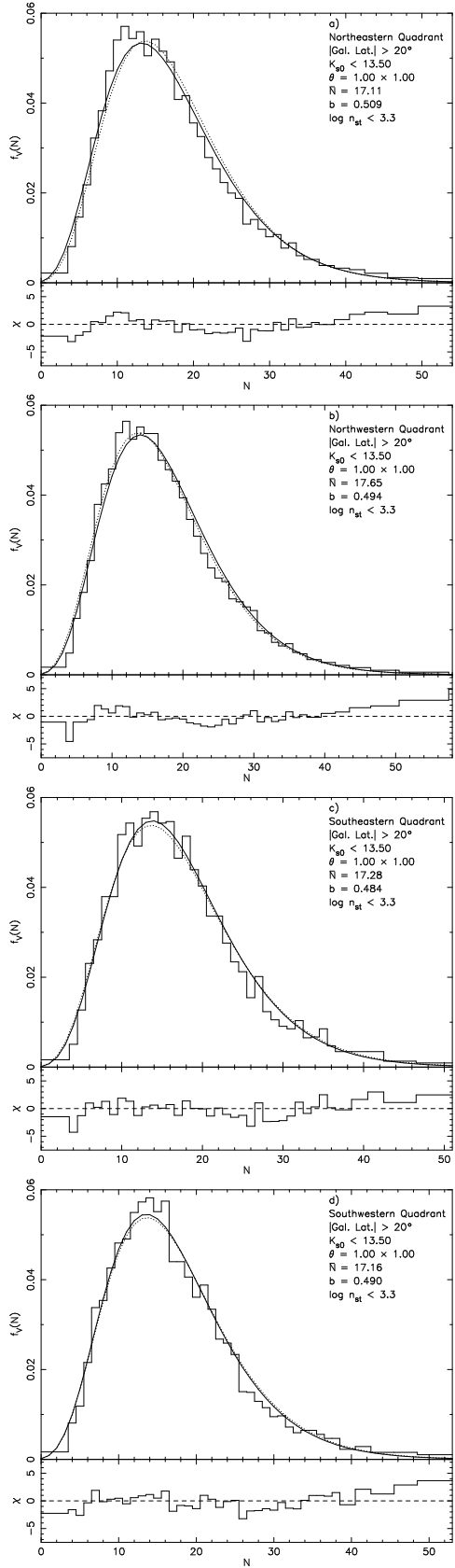

\epsfig{file=f4a.ps,angle=-90,clip=,width=2.5in}\\
\epsfig{file=f4b.ps,angle=-90,clip=,width=2.5in}\\
\epsfig{file=f4c.ps,angle=-90,clip=,width=2.5in}\\
\epsfig{file=f4d.ps,angle=-90,clip=,width=2.5in}
\caption{
As Figure~\ref{fig:GQED_size}, except the figures indicate different
quadrants all with $1\degr$ cells on the sky. The dotted lines
indicate the best-fit model from the full sample with values of $b$
and $\overline{N}$ given in Table~\ref{tbl:GQED}.
\label{fig:GQED_quad}}
\end{figure}

The 2MASS sample is the first that is large enough for a precise
determination of variations among different parts of the
sky. Figures~\ref{fig:GQED_quad}a - d give our results for dividing
the sky into four independent quadrants.
Each quadrant
contains between about 5,500 and 6,200 usable cells. Fluctuations
around equation~(\ref{eq:GQED}) are very small, though naturally they
differ in detail. Among these quadrants the variance of $\overline{N}$
is 1.2\% and of $b$ is just 1.9\%. For these quadrants divided into
$4\degr \times 4\degr$ cells, for which each quadrant has between about
250 and 330 cells, the variance of $\overline{N}$ is 1.5\% and of $b$
is 2.6\%. The smaller fluctuations among the larger cells, which are
more representative of the total galaxy distribution, partly
compensate for the smaller total number of such cells. This suggests
that a very conservative estimate of the uncertainty in $b$ is 3\%. By
averaging the two quadrants with the lowest values of $b$ to give
0.485 and comparing with the averages of the two quadrants giving the
largest value of $b$ = 0.494, we would obtain an estimate of 2\% for
the uncertainty of $b$ in cells of one square degree.
For comparison, earlier estimates of $b$ from the catalogs mentioned
in Section~\ref{sec:intro_2MASS_dist} gave uncertainties of 5-10\% for
$b$.

It is interesting to see the effects of a large cluster, in particular
the Shapley Supercluster, on these statistics. On small scales it is
obviously noticeable, as pointed out in the northwest quadrant of
Figure~\ref{fig:GQED_allsky}. If it were important when averaged over
the whole sky, or even over a quadrant, we would consider it a
sufficient anomaly to question whether it could have formed by
gravitational many-body clustering alone, or whether it needed to
arise from large perturbations in the early universe.

\begin{figure}
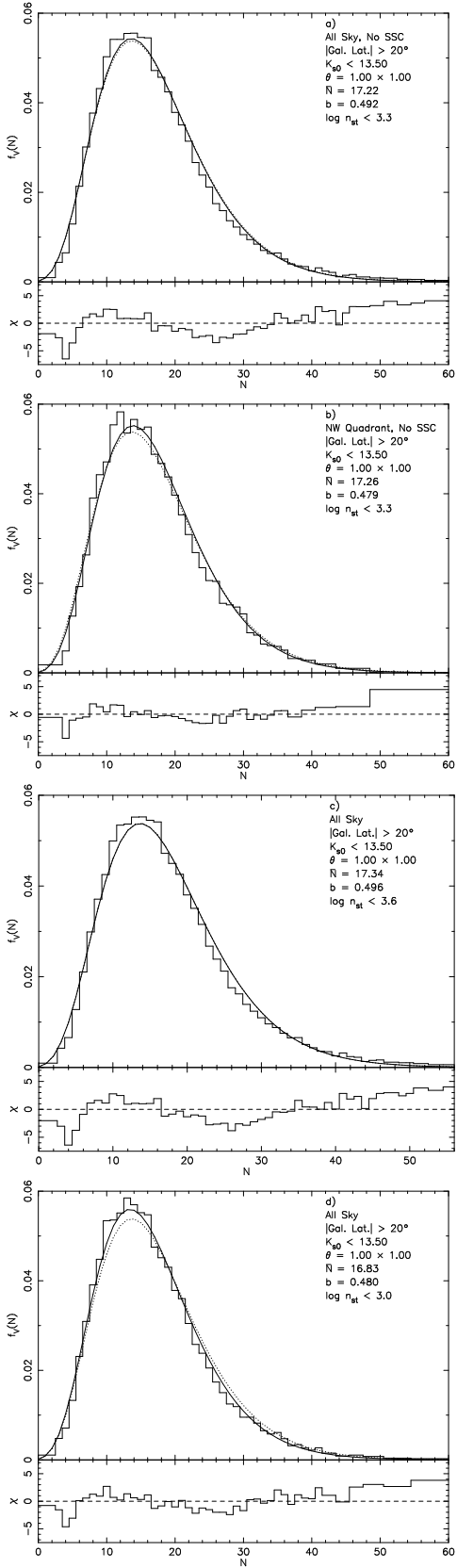

\epsfig{file=f5a.ps,angle=-90,clip=,width=2.5in}\\
\epsfig{file=f5b.ps,angle=-90,clip=,width=2.5in}\\
\epsfig{file=f5c.ps,angle=-90,clip=,width=2.5in}\\
\epsfig{file=f5d.ps,angle=-90,clip=,width=2.5in}
\caption{
As Figure~\ref{fig:GQED_size}, except (a) and (b) indicate removing
the Shapley Supercluster from the entire sky and the northwestern
quadrant, while (c) and (d) indicate changing the stellar density at
which cells are excluded, all with $1\degr$ size cells on the sky.
The dotted lines indicate the best-fit model from the full sample
with values of $b$ and $\overline{N}$ given in Table~\ref{tbl:GQED}.
\label{fig:GQED_other}}
\end{figure}

Figures~\ref{fig:GQED_other}a, b show the effects of removing the
Shapley Supercluster from the all sky sample and from the northwest
quadrant.
This is effected by removing all cells within 10 degrees of
$(\alpha_{gal},\delta_{gal}) = (312.457\degr,30.754\degr)$.
Comparison with the same all sky samples in
Figures~\ref{fig:GQED_size}c, d shows that removing the SSC reduces
$\overline{N}$ slightly, as expected, but reduces the value of $b$ by
less than 1\%, and changes the histograms by less than the effects of
the bin shift. On the smaller scale of a quadrant, however, there is
less dilution and the effects of the SSC are more noticeable.
The
value of $b$ is now reduced by 3\%, compared with
Figure~\ref{fig:GQED_quad}b, although the histograms remain very
similar.

A small fraction of anomalies in the 2MASS Galaxy (extended source)
Catalog may be produced by contamination from point sources so close
to each other that they masquerade as an extended source. For this to
happen, their colors and magnitudes would also have to coincide with
our criteria described in Section~\ref{sec:analysis_2MASS_dist} for a
galaxy.
Figures~\ref{fig:GQED_size}c and \ref{fig:GQED_other}c, d
illustrate this effect.
In Figure~\ref{fig:GQED_other}c we employ the less stringent criterion
that all cells contain less than $10^{3.6}$ point sources, compared to
limits of $10^{3.3}$ and $10^{3.0}$ in Figures~\ref{fig:GQED_size}c
and \ref{fig:GQED_other}d. This less stringent criterion induces
slight fluctuations in the histograms but does not change the values
of $\overline{N}$ or $b$ to three significant figures. The more
stringent criterion of Figure~\ref{fig:GQED_other}d reduces both the
number of cells and galaxies considerably (see Table~\ref{tbl:GQED}
below) and has a 3\% smaller value of b. This, and examination of
Table~\ref{tbl:GQED}, suggest that the limit of $10^{3.3}$ is a
reasonable compromise.

\begin{figure}
\epsfig{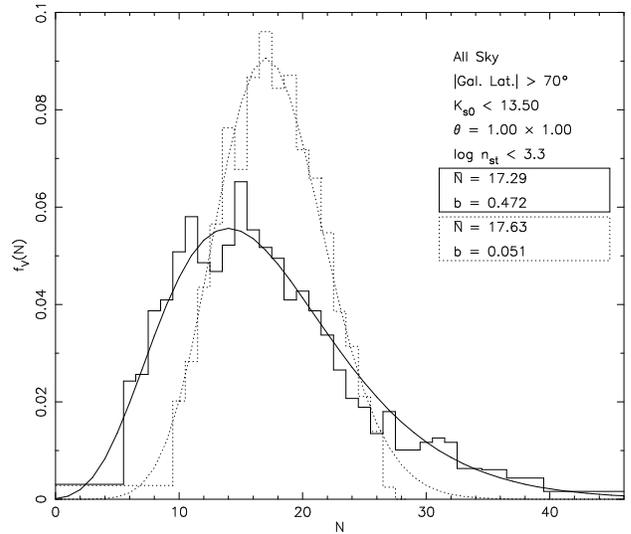}
\caption{
Least-squares fits (continuous lines) of $f_{V}(N)$ to the observed
histograms for likely galaxies (2MASS XSC, solid lines) and likely stars
(2MASS PSC, dotted lines).
\label{fig:GQED_star}}
\end{figure}

In Figure~\ref{fig:GQED_star}, we directly compare a sample from the
2MASS Point Source Catalog (PSC, mainly stars) to a sample from the
2MASS XSC (mainly galaxies). To mitigate the effects of the Galactic
disc, the comparison uses a Galactic latitude cutoff of $70 \degr$. In
the PSC, we first selected objects with valid photometry in all three
bands, where $K_{s,0} < 13.5$. Since there are many more objects in
the PSC, we then randomly selected objects with a probability that
would give approximately the same $\overline{N}$ as seen in the same
area as the XSC.  We also used the same stellar density
requirements. The two distributions (XSC - solid line, PSC - dotted
line) are clearly different with the PSC showing much less clustering
(lower $b$).  We also found that reducing the $\overline{N}$ for the PSC
by magnitude selection produced a similar value of $b \approx 0.05$.

\begin{figure*}
\plotone{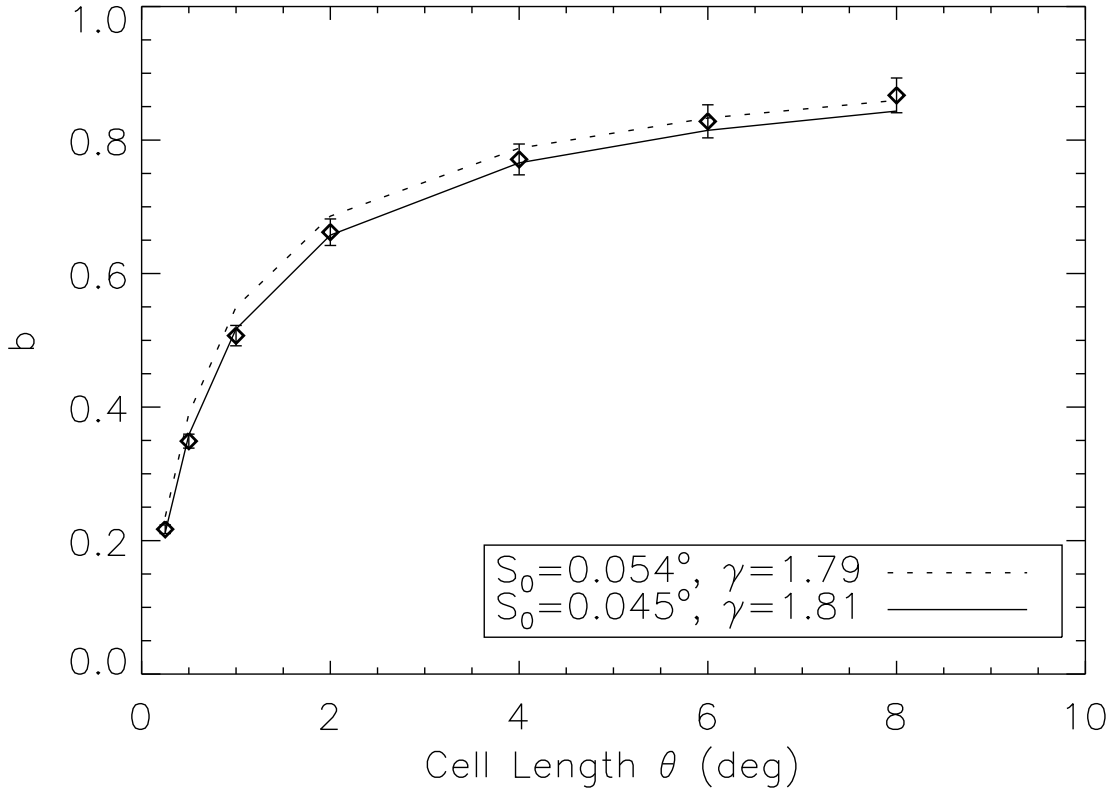}
\caption{
Plot of $b(\theta)$. The error bars represent conservative
uncertainties of $\pm 3$\% for each value of $b$.
The predicted $b(\theta)$ (equation~\ref{eq:b_theta})
for the two-galaxy angular correlation function as measured
for 2MASS in \citet{MMK+_2005} (dotted-line) and in this paper (solid-line).
\label{fig:b_theta}}
\end{figure*}

The dependence of $b$ on the scale and shape of the cells incorporates
important information about the two-galaxy correlation function, as
indicated by equation~(\ref{eq:b}). Through the form of
equation~(\ref{eq:GQED}), this dependence of $b$ also incorporates
much information about higher order correlations. This scale
dependence can be used to test the applicability of computer
simulations. Indeed a wide range of cold dark matter models do not
agree with the scale dependence of $b$ in the IRAS Catalog
\citep{SMS_1994}. The increased precision of the 2MASS Catalog now
makes $b(\theta)$ an even more stringent test of computer
simulations. The $b(\theta)$ dependence is shown in
Figure~\ref{fig:b_theta}, where we have indicated conservative
uncertainties of $\pm 3\%$ for each value of $b$.
Typical CDM models,
especially those whose initial perturbation spectra produce large dark
matter haloes that contain many galaxies, have too many voids and too
many filaments at the present time (Baertschiger \& Sylos-Labini 2004,
private communication).
Even if they agree approximately with the functional form of
equation~(\ref{eq:GQED}), their excessive voids and
filamentary structures would give significantly greater values of $b$
on small or large scales than the observations in Figure~\ref{fig:b_theta}
show.
In Figure~\ref{fig:b_theta}, we also compare $b(\theta)$ with
values based on the two-galaxy angular correlation function, $W(s)$.
\citet{LS_1992} derived
$b(\theta)$ for square cells of size $\omega = \theta \times \theta$ square
degrees
\begin{equation}
  \label{eq:b_theta}
  b(\theta) =  1-(1+\overline{N} \ s_{0}^{\gamma-1} \ 
                        C_{\gamma} \ \theta^{5-\gamma})^{-0.5},
\end{equation}
where $W(s) = (s/s_{0})^{1-\gamma}$, $\overline{N}$ is the expected
number of galaxies projected into a cell, and
$C_{\gamma}$ is a coefficient evaluated numerically from
$J \equiv \int \! \! \int_{\omega} \theta^{1-\gamma} \ 
d \omega_{1} \ d \omega_{2} \ = C_{\gamma} \ \theta^{5-\gamma}$
as in \citet{TK_1969} giving $C_\gamma=2.25$ for
$\gamma =1.8$. \citet{MMK+_2005} find $\gamma=1.79 \pm 0.02$
and $s_0 = 0.054 \pm 0.008$ for the 2MASS catalog at
$K_{s,0} < 13.5$, from which we derive the dotted line in
Figure~\ref{fig:b_theta}. This 
fit is less good at smaller $\theta$ and has a $\chi^{2} = 33.6$ for 5 degrees
of freedom. Fitting $\gamma$ and $s_{0}$ from $b(\theta)$,
we find $\gamma=1.81 \pm 0.04$
and $s_0 = 0.045 \pm 0.003$ (solid line in
Figure~\ref{fig:b_theta}), with a $\chi^{2} = 3.0$, overlapping with
the values from \citet{MMK+_2005}. The distribution function,
equation~(\ref{eq:GQED}), therefore provides a good method for
measuring the angular correlation function.

Table~\ref{tbl:GQED} below gives a more detailed summary of results
for a wider range of analyses than
Figures~\ref{fig:GQED_size}--\ref{fig:b_theta}.
The first section of
Table~\ref{tbl:GQED} examines samples with different
stellar density limits in $1 \times 1 \degr$ cells over all the
unobscured sky above $\pm 20\degr$ Galactic latitude but with
different magnitude cutoffs. Values of $\overline{N}$ and $b$ are
determined in three different ways. In the first column of values,
$\overline{N}$ is found exactly by averaging its value over all cells
of each sample, and $b$ is found from the variance of counts in these
cells using equation~(\ref{eq:var}). In the second column,
$\overline{N}$ is found as before but $b$ is found using a
least-squares fit to equation~(\ref{eq:GQED}). In the third column,
$\overline{N}$ and $b$ are both obtained from a two-parameter
least-squares fit to equation~(\ref{eq:GQED}). In all cases, the fits
are visually as good as those illustrated in
Figures~\ref{fig:GQED_size}--\ref{fig:GQED_other}.
We have also given
values of $\chi^{2}$ per degree of freedom, although as mentioned in
Section~\ref{sec:analysis_2MASS_dist} these values cannot be used to
calculate absolute or relative probabilities for comparisons among
themselves or with other distributions. Moreover, these $\chi^{2}$
values tend to be dominated by the large $N$ tail of the
distributions. The values of $b$ do not depend significantly on
whether or not $\overline{N}$ is a fitting parameter or is directly
determined from the counts in cells. The values of $b$ determined from
the variance using equation~(\ref{eq:var}) tend to be high for small
scales but agree well with the values from the fits for larger cells
which are more representatiave (even though there are fewer of them).
The remaining sections of Table~\ref{tbl:GQED} examine the effects of
varying Galactic latitude cutoffs, varying the size of the cell, and
varying the regions of the sky to which the GQED is fit.

\begin{figure*}
\plotone{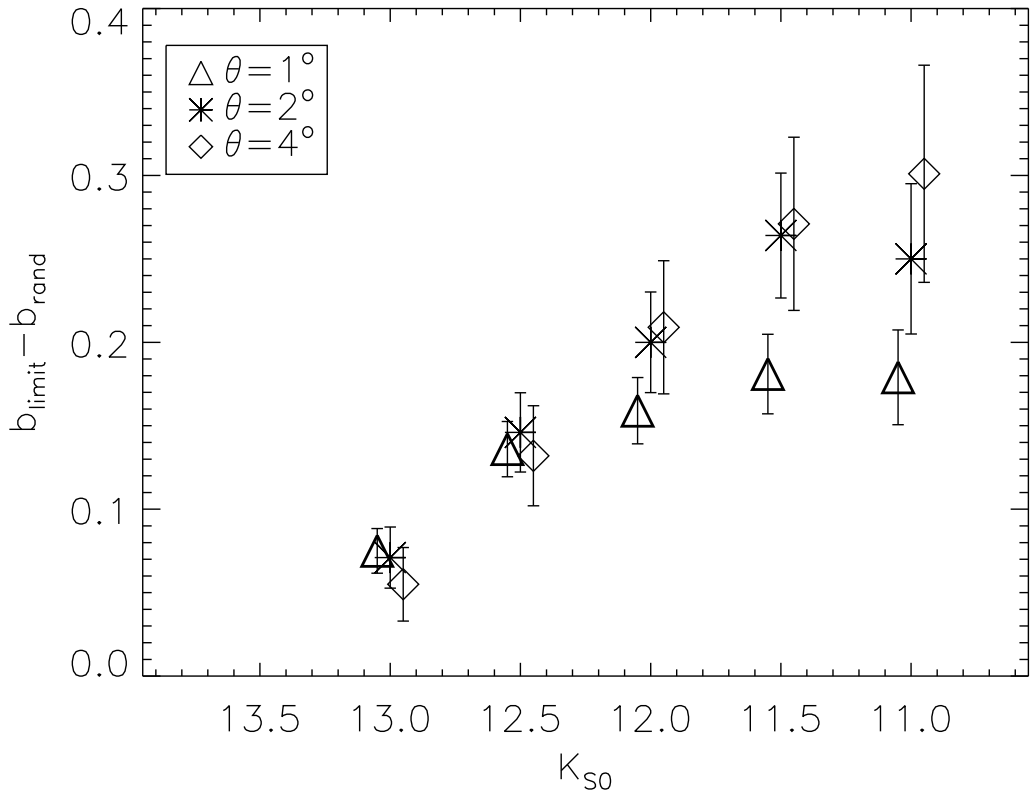}
\caption{
Plot of the differences between values of $b$ measured from magnitude limited
subsamples and from randomly selected subsamples of equal size as a function of
magnitude for three cell sizes.
Small shifts in plotted magnitude are
made so that individual data-points and error bars can be distinguished.
The error bars represent conservative uncertainties scaled by
$\overline{N}$ to $\pm 3$\% of $b$ for $K_{s,0}=13.5$.
\label{fig:b_rand_mag}}
\end{figure*}

In Table~\ref{tbl:GQED_rand}, we compare the effects of sampling the
galaxy catalog by magnitude cuts with sampling by a random selection
that yields approximately the same number of galaxies at each
magnitude cut.
For these comparisons, we have used the usual stellar
density cutoff ($n_{st} < 10^{3.3}$), All-sky region, and Galactic latitude
cutoff (above $\pm 20\degr$).
The value of $b$ in a subsample selected
randomly with a probability $p$ is related to the value of the parent
sample by (e.g., \citealt{S_2000})
\begin{equation}
  \label{eq:random}
(1-b_{subsample})^2 = \frac{(1-b_{parent})^2 }
                           {1-(1-p)(2-b_{parent})b_{parent}}.
\end{equation}
The values of $b$ in Table~\ref{tbl:GQED_rand} for the random
sampling of the 13.5 magnitude limited galaxy catalog compared
with the values of $b$ in Table~\ref{tbl:GQED} confirm
equation~(\ref{eq:random}).
The pairs of rows in Table~\ref{tbl:GQED_rand} give observed parameters
for a magnitude cut and for a corresponding Poisson selected subsample.
This analysis is for cells of $1 \times 1
\degr$, $2 \times 2 \degr$, and $4 \times 4 \degr$.
Figure~\ref{fig:b_rand_mag} shows the differences
between $b$ measured for the magnitude cut and $b$ for an
equivalent randomly selected subsample.

\begin{figure*}
\plotone{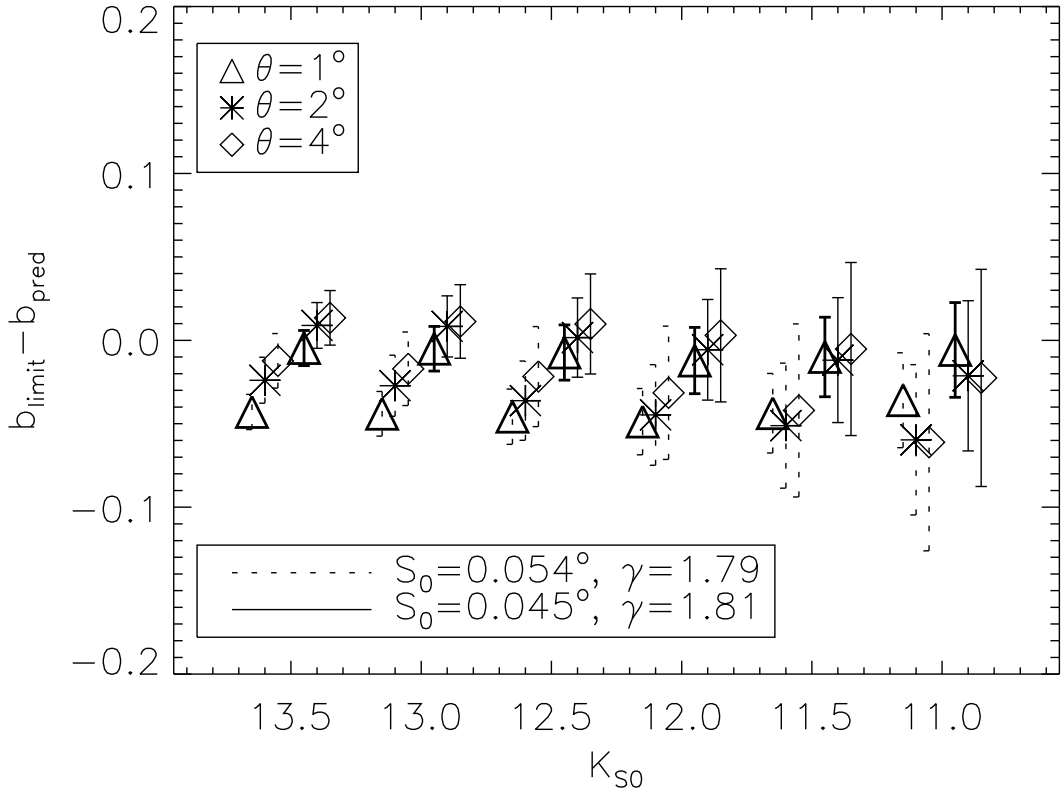}
\caption{
Plot of the differences between values of $b$ measured from magnitude limited
subsamples and predictions for $b$ from equations \ref{eq:b_theta} and
\ref{eq:theta_mag} as a function of magnitude for three cell sizes.
We use the two-galaxy angular correlation function as measured
for 2MASS in \citet{MMK+_2005} (dotted error-bars) and in this paper
(solid error-bars).
Small shifts in plotted magnitude are
made so that individual data-points and error bars can be distinguished.
The error bars represent conservative uncertainties scaled by
$\overline{N}$ to $\pm 3$\% of $b$ for $K_{s,0}=13.5$.
\label{fig:b_pred_mag}}
\end{figure*}

The magnitude selected subsample always has a higher observed value of
$b$ than the corresponding randomly selected subsample.
This effect appears to be mainly geometric.  As one goes to
brighter subsamples, their average volume decreases provided the
distribution is statistically homogeneous.  This affects $W(s)$ by
increasing $s_{0}$.  Assuming the average radial distance sampled follows
$D(m) \propto 10^{m/5}$, and with the usual scaling argument,
$W(s) \propto W(s \ D(m))/D(m)$ \citep{P_1993},
\begin{equation}
  \label{eq:theta_mag}
  s_{0,m_2} = s_{0,m_1} 10^{[(m_1-m_2)/5] [\gamma/(\gamma-1)]}.
\end{equation}
Combining equations \ref{eq:b_theta} and \ref{eq:theta_mag}, we can
predict $b(\theta, m)$ based on $W(s)$ at $K_{s,0}=13.5.$ In
Figure~\ref{fig:b_pred_mag}, we show the measured $b$ minus this
predicted $b$. Using the angular correlation, $W(s, 13.5)$,
from \citet{MMK+_2005},
there are some small discrepancies for small $\theta$. With our
best-fit $W(s, 13.5)$, there are no discrepancies.

\section{Summary and Discussion}
\label{sec:discussion_2MASS_dist}

Our main result is that at a level of precision 2-3 times better than
possible with previous catalogs, the detailed form of the galaxy
distribution in the 2MASS Catalog agrees with its prediction from
gravitational cosmological many-body theory.
The value of the clustering parameter $b$ in equation~(\ref{eq:b})
appears to be $b = 0.867\pm0.026$ on large scales in the present
universe with $z \lesssim 0.1$. This is in accordance with the value of
$0.83\pm0.05$ for the much smaller but also highly complete
three-dimensional Pisces-Perseus sample
\citep{SH_1998}. Table~\ref{tbl:GQED} shows that as the cell size
increases the values of $b$ deduced from equation~(\ref{eq:var}) agree
better with the values in the $\chi^{2}$ fits. This is essentially
because larger cells are more representative of the global clustering,
and their variance is more accurately determined. These values of $b$
also agree with N-body simulations at the present time in $\Omega_0 =
1$ models (summarized in \citealt{S_2000}).
Agreement with gravitational many-body theory need not have occurred
at this level, especially since there are at least four major
processes that could have destroyed it. The reasons that the agreement
survives may be quite instructive.

First, a non-uniform distribution of dark gravitating matter that
differs from the luminous galaxy distribution could significantly
modify the observed galaxy distribution, in disagreement with the form
of equation~(\ref{eq:GQED}). If dark haloes contain many galaxies, the
resulting voids and filaments could exceed those observed. The value
of $b$ in such models would be too high. Moreover models with
overlapping individual dark matter haloes would soften the
gravitational potential significantly and their resulting peculiar
velocity distribution function would also disagree with observations
\citep{LS_2004}. All this indicates that when galaxies cluster
most of the dark matter is located in the haloes of individual
galaxies.

Second, if the initial perturbations that stimulated galaxy clustering
have survived into the present non-linear regime, they could produce
departures from the form of equation~(\ref{eq:GQED}). This would be
especially true if the initial conditions were strongly correlated or
anti-correlated over large scales and have not had time to relax
through a sequence of quasi-equilibrium states \citep{S_2000}.
For example, an initial power law perturbation spectrum would have to
have an exponent between $-1 \lesssim n \lesssim 1$ for such
relaxation \citep{I_1990}). Of course the initial power spectrum may
not have been a simple power law; then this question becomes much more
complicated and has not yet been systematically explored.

Third, there is considerable evidence, especially from the Hubble Deep
Fields, that merging has played a significant role in galaxy formation
which extends into the present. Merging tends to modify the form of
$f_{V}(N)$ slightly and decrease the value of $b$ \citep{FS_1997}. The
effects of merger induced starbursts modify the luminosity function,
and complicate the selection of galaxies in the distribution function,
probably depending on the local environment.

Evidently mergers do not destroy the distribution functions in
Section~\ref{sec:results_2MASS_dist}. There may be several reasons for
this. Although individual mergers may be dramatic, only a small
percentage of galaxies merge at any given time so they do not have
much effect on the overall statistics. When galaxies merge, they
generally do so near the center of mass with their nearest
neighbor. These centers of mass may have a distribution which
approximately follows that of the galaxies themselves. Moreover
intervals between mergers are often long compared with the local
orbital relaxation time, so the merged galaxy has time to acquire the
quasi-equilibrium distribution of equation~(\ref{eq:GQED}).

Fourth, although the distribution of equation~(\ref{eq:GQED}) is for
systems in which all galaxies have the same mass, N-body simulations
(e.g., \citealt{IIS_1993}) show that a range of masses generally
enhances clustering and increases the value of $b$ compared with
single mass cases. The effect is relatively small, typically $\sim$10\%,
because much of the later clustering involves the collective
interaction of individual galaxies with groups of galaxies, or of
groups with other groups, for which the masses of individual galaxies
are less relevant. For a wide range of masses, there is a tendency for
massive galaxies to acquire less massive galaxies as satellites. How
much this affects the distribution function depends on the relative
total numbers of low and high mass galaxies and, for the observations,
on detailed models of mass-luminosity galaxy selection.

While it is conceivable that these four, and perhaps other, effects
might conspire to leave the distribution function of
equation~(\ref{eq:GQED}) invariant, it seems more likely to us that
each effect is individually small. Nonetheless they may be present to
some degree since the fits to equation~(\ref{eq:GQED}) are not
perfect. In particular, there seem to be slight systematic tendencies
for the observed peaks to exceed the theoretical peaks, for a small
observed excess in the large $N$ tail, and for a mild observed deficit
between the peak and this tail. This holds for many samples, but not
all. It seems somewhat more prominent in samples with greater stellar
contamination, but might also be produced by some of the effects
mentioned previously. At present, it seems that tracking down these
small discrepancies will require a combination of even more extensive
observations and more detailed models.

We have also found (Table~\ref{tbl:GQED_rand}) that values of $b$ for
subsamples limited by brighter apparent magnitude cutoffs are higher
than $b$ would be for a subsample selected randomly with the same
probability. This was not evident in an earlier analysis
\citep{SC_1991} of the Zwicky catalog, probably because this earlier
catalog contains only about 7\% as many galaxies as 2MASS and
differences of $\le 1 \, {\rm mag}$ were explored. This effect can be
explained almost entirely by the smaller volume of subsamples with
brighter magnitude cutoffs, as Figure~\ref{fig:b_pred_mag} shows. From
equation~(\ref{eq:b}), the spatially integrated effect of luminosity
clustering as observed in the luminosity dependence of the two-point
correlation function (e.g., \citealt{MMB_1994,BMD+_1996,NBH+_2001}) is
therefore small. Similarly, the good fits of our magnitude limited
subsamples to equation~(\ref{eq:GQED}), which includes all the higher
order correlation functions and moments of the distribution, show that
luminosity clustering is small when averaged over large statistically
homogeneous volumes containing many clusters. Physically, this would be
expected if most of the clustering were produced by collective
gravitational interactions (e.g., \citealt{S_2000}).

With further information about other properties of the 2MASS
galaxies, distribution functions will make possible new insights into
relative clusterings based on morphological types, colors, presence
of satellite and tidally interacting galaxies, etc. This will contribute
to a wealth of constraints on galaxy formation and evolution in more
detailed models.

\acknowledgements

This publication makes use of data products from the Two Micron All
Sky Survey, which is a joint project of the University of
Massachusetts and the Infrared Processing and Analysis
Center/California Institute of Technology, funded by the National
Aeronautics and Space Administration and the National Science
Foundation. We are especially grateful to Mike Skrutskie for
helpful advice about the 2MASS survey
and to the anonymous referee for especially helpful comments.
GRS acknowledges the receipt
of an Achievement Reward for College Scientists fellowship. Partial
support was also provided by the F.H. Levinson Fund of the Peninsula
Community Foundation.

\pagestyle{empty}
\clearpage
\setlength{\tabcolsep}{0.05in}
\LongTables 
\begin{landscape}
\begin{deluxetable}{cccccccccccccccccc}
\tablewidth{0pt}
\tabletypesize{\footnotesize}
\tablecaption{GQED Fits to 2MASS Data\label{tbl:GQED}}
\tablehead{
Max&
Max&
&
Min&
&
Total&
Total&
\multicolumn{3}{c}{From Variance}&
&
\multicolumn{3}{c}{Fitting $b$}&
&
\multicolumn{3}{c}{Fitting $\overline{N}$ \& $b$}
\\
\cline{8-10} \cline{12-14}  \cline{16-18}\\[-2ex]
\colhead{$K_{s0}$}&
\colhead{$log(n_{st})$}&
\colhead{Region}&
\colhead{$\vert \delta_{Gal} \vert$ }&
\colhead{$\theta$}&
\colhead{Galaxies}&
\colhead{Cells}&
\colhead{$\overline{N}$}&
\colhead{$b$}&
\colhead{$\chi^{2}$/Dof}&
&
\colhead{$\overline{N}$}&
\colhead{$b$}&
\colhead{$\chi^{2}$/Dof}&
&
\colhead{$\overline{N}$}&
\colhead{$b$}&
\colhead{$\chi^{2}$/Dof}\\
&
}
\startdata
\multicolumn{18}{l}{For Varying $n_{st}$, $K_{s0}$ Cutoffs}\\
$11.00$\phm{$^{a}$}     &$3.0$&All                 &$20$&$1.00$                 &$\phn\phn6717$&$      \phn15199$&$\phn\phn\phn0.44$&$0.228$&$   \phn\phn26.4/\phn7$&$$&$\phn\phn\phn0.44$&$0.203$&$   \phn14.9/\phn6$&$$&$\phn\phn\phn0.44$&$0.200$&$   \phn14.2/\phn5$\\
$11.00$                 &$3.3$&All                 &$20$&$1.00$                 &$   \phn10940$&$      \phn23348$&$\phn\phn\phn0.47$&$0.264$&$   \phn\phn84.5/\phn8$&$$&$\phn\phn\phn0.47$&$0.223$&$   \phn37.6/\phn7$&$$&$\phn\phn\phn0.46$&$0.218$&$   \phn34.8/\phn6$\\
$11.00$                 &$3.6$&All                 &$20$&$1.00$                 &$   \phn11421$&$      \phn24171$&$\phn\phn\phn0.47$&$0.274$&$      \phn116.0/\phn8$&$$&$\phn\phn\phn0.47$&$0.223$&$   \phn42.8/\phn7$&$$&$\phn\phn\phn0.46$&$0.217$&$   \phn39.0/\phn6$\\
$11.50$                 &$3.0$&All                 &$20$&$1.00$                 &$   \phn13655$&$      \phn15199$&$\phn\phn\phn0.90$&$0.283$&$   \phn\phn55.9/\phn9$&$$&$\phn\phn\phn0.90$&$0.248$&$   \phn23.9/\phn8$&$$&$\phn\phn\phn0.89$&$0.245$&$   \phn22.3/\phn7$\\
$11.50$                 &$3.3$&All                 &$20$&$1.00$                 &$   \phn22252$&$      \phn23348$&$\phn\phn\phn0.95$&$0.329$&$      \phn250.8/   12$&$$&$\phn\phn\phn0.95$&$0.266$&$   \phn87.3/   11$&$$&$\phn\phn\phn0.92$&$0.257$&$   \phn76.7/   10$\\
$11.50$                 &$3.6$&All                 &$20$&$1.00$                 &$   \phn23131$&$      \phn24171$&$\phn\phn\phn0.96$&$0.333$&$      \phn288.8/   12$&$$&$\phn\phn\phn0.96$&$0.265$&$   \phn91.7/   11$&$$&$\phn\phn\phn0.93$&$0.256$&$   \phn79.3/   10$\\
$12.00$                 &$3.0$&All                 &$20$&$1.00$                 &$   \phn27872$&$      \phn15199$&$\phn\phn\phn1.83$&$0.340$&$      \phn158.5/   13$&$$&$\phn\phn\phn1.83$&$0.291$&$   \phn72.9/   12$&$$&$\phn\phn\phn1.79$&$0.284$&$   \phn66.6/   11$\\
$12.00$                 &$3.3$&All                 &$20$&$1.00$                 &$   \phn45385$&$      \phn23348$&$\phn\phn\phn1.94$&$0.388$&$      \phn479.2/   17$&$$&$\phn\phn\phn1.94$&$0.317$&$      166.8/   16$&$$&$\phn\phn\phn1.88$&$0.305$&$      144.5/   15$\\
$12.00$                 &$3.6$&All                 &$20$&$1.00$                 &$   \phn47145$&$      \phn24171$&$\phn\phn\phn1.95$&$0.389$&$      \phn505.4/   17$&$$&$\phn\phn\phn1.95$&$0.317$&$      169.6/   16$&$$&$\phn\phn\phn1.88$&$0.305$&$      145.6/   15$\\
$12.50$                 &$3.0$&All                 &$20$&$1.00$                 &$   \phn58035$&$      \phn15199$&$\phn\phn\phn3.82$&$0.398$&$      \phn193.0/   20$&$$&$\phn\phn\phn3.82$&$0.355$&$   \phn94.8/   19$&$$&$\phn\phn\phn3.75$&$0.349$&$   \phn88.0/   18$\\
$12.50$                 &$3.3$&All                 &$20$&$1.00$                 &$   \phn94467$&$      \phn23348$&$\phn\phn\phn4.05$&$0.453$&$      \phn731.8/   24$&$$&$\phn\phn\phn4.05$&$0.382$&$      243.3/   23$&$$&$\phn\phn\phn3.92$&$0.368$&$      209.6/   22$\\
$12.50$                 &$3.6$&All                 &$20$&$1.00$                 &$   \phn97926$&$      \phn24171$&$\phn\phn\phn4.05$&$0.453$&$      \phn751.8/   24$&$$&$\phn\phn\phn4.05$&$0.382$&$      247.3/   23$&$$&$\phn\phn\phn3.92$&$0.369$&$      212.5/   22$\\
$12.75$                 &$3.0$&All                 &$20$&$1.00$                 &$   \phn84186$&$      \phn15199$&$\phn\phn\phn5.54$&$0.432$&$      \phn204.7/   25$&$$&$\phn\phn\phn5.54$&$0.392$&$   \phn98.8/   24$&$$&$\phn\phn\phn5.46$&$0.386$&$   \phn92.7/   23$\\
$12.75$                 &$3.3$&All                 &$20$&$1.00$                 &$      136610$&$      \phn23348$&$\phn\phn\phn5.85$&$0.485$&$      \phn836.6/   29$&$$&$\phn\phn\phn5.85$&$0.417$&$      264.1/   28$&$$&$\phn\phn\phn5.69$&$0.405$&$      232.3/   27$\\
$12.75$                 &$3.6$&All                 &$20$&$1.00$                 &$      141599$&$      \phn24171$&$\phn\phn\phn5.86$&$0.485$&$      \phn851.7/   29$&$$&$\phn\phn\phn5.86$&$0.417$&$      268.0/   28$&$$&$\phn\phn\phn5.70$&$0.406$&$      235.6/   27$\\
$13.00$                 &$3.0$&All                 &$20$&$1.00$                 &$      122608$&$      \phn15199$&$\phn\phn\phn8.07$&$0.461$&$      \phn271.4/   31$&$$&$\phn\phn\phn8.07$&$0.422$&$      143.9/   30$&$$&$\phn\phn\phn7.95$&$0.416$&$      136.1/   29$\\
$13.00$                 &$3.3$&All                 &$20$&$1.00$                 &$      197738$&$      \phn23348$&$\phn\phn\phn8.47$&$0.514$&$         1108.0/   37$&$$&$\phn\phn\phn8.47$&$0.446$&$      364.2/   36$&$$&$\phn\phn\phn8.24$&$0.433$&$      326.3/   35$\\
$13.00$                 &$3.6$&All                 &$20$&$1.00$                 &$      204826$&$      \phn24171$&$\phn\phn\phn8.47$&$0.514$&$         1111.1/   37$&$$&$\phn\phn\phn8.47$&$0.447$&$      366.9/   36$&$$&$\phn\phn\phn8.25$&$0.434$&$      328.1/   35$\\
$13.25$                 &$3.0$&All                 &$20$&$1.00$                 &$      178009$&$      \phn15199$&$   \phn\phn11.71$&$0.492$&$      \phn264.8/   38$&$$&$   \phn\phn11.71$&$0.456$&$      138.3/   37$&$$&$   \phn\phn11.55$&$0.450$&$      129.6/   36$\\
$13.25$                 &$3.3$&All                 &$20$&$1.00$                 &$      285786$&$      \phn23348$&$   \phn\phn12.24$&$0.540$&$         1073.2/   45$&$$&$   \phn\phn12.24$&$0.479$&$      359.9/   44$&$$&$   \phn\phn11.94$&$0.467$&$      320.5/   43$\\
$13.25$                 &$3.6$&All                 &$20$&$1.00$                 &$      296039$&$      \phn24171$&$   \phn\phn12.25$&$0.540$&$         1082.8/   46$&$$&$   \phn\phn12.25$&$0.479$&$      369.2/   45$&$$&$   \phn\phn11.95$&$0.467$&$      328.2/   44$\\
$13.50$                 &$3.0$&All                 &$20$&$1.00$                 &$      258815$&$      \phn15199$&$   \phn\phn17.03$&$0.520$&$      \phn308.5/   47$&$$&$   \phn\phn17.03$&$0.485$&$      168.0/   46$&$$&$   \phn\phn16.83$&$0.480$&$      159.9/   45$\\
$13.50$                 &$3.1$&All                 &$20$&$1.00$                 &$      333937$&$      \phn19170$&$   \phn\phn17.42$&$0.541$&$      \phn617.0/   52$&$$&$   \phn\phn17.42$&$0.495$&$      275.6/   51$&$$&$   \phn\phn17.14$&$0.487$&$      256.9/   50$\\
$13.50$                 &$3.2$&All                 &$20$&$1.00$                 &$      382096$&$      \phn21652$&$   \phn\phn17.65$&$0.554$&$      \phn831.0/   54$&$$&$   \phn\phn17.65$&$0.504$&$      346.7/   53$&$$&$   \phn\phn17.32$&$0.495$&$      318.9/   52$\\
$13.50$\tablenotemark{a}&$3.3$&All                 &$20$&$1.00$                 &$      413339$&$      \phn23348$&$   \phn\phn17.70$&$0.565$&$         1126.1/   54$&$$&$   \phn\phn17.70$&$0.507$&$      397.7/   53$&$$&$   \phn\phn17.34$&$0.496$&$      359.9/   52$\\
$13.50$                 &$3.4$&All                 &$20$&$1.00$                 &$      422901$&$      \phn23894$&$   \phn\phn17.70$&$0.564$&$         1130.7/   55$&$$&$   \phn\phn17.70$&$0.507$&$      404.5/   54$&$$&$   \phn\phn17.33$&$0.497$&$      365.9/   53$\\
$13.50$                 &$3.5$&All                 &$20$&$1.00$                 &$      427446$&$      \phn24136$&$   \phn\phn17.71$&$0.563$&$         1142.7/   56$&$$&$   \phn\phn17.71$&$0.507$&$      408.4/   55$&$$&$   \phn\phn17.34$&$0.496$&$      369.0/   54$\\
$13.50$                 &$3.6$&All                 &$20$&$1.00$                 &$      428109$&$      \phn24171$&$   \phn\phn17.71$&$0.563$&$         1137.5/   56$&$$&$   \phn\phn17.71$&$0.507$&$      406.5/   55$&$$&$   \phn\phn17.34$&$0.496$&$      367.2/   54$\\
$13.50$                 &$4.0$&All                 &$20$&$1.00$                 &$      428578$&$      \phn24194$&$   \phn\phn17.71$&$0.563$&$         1142.0/   56$&$$&$   \phn\phn17.71$&$0.507$&$      408.1/   55$&$$&$   \phn\phn17.34$&$0.497$&$      368.5/   54$\\
\multicolumn{18}{l}{For Varying Galactic Latitude Cutoffs}\\
$13.50$\tablenotemark{a}&$3.3$&All                 &$20$&$1.00$                 &$      413339$&$      \phn23348$&$   \phn\phn17.70$&$0.565$&$         1126.1/   54$&$$&$   \phn\phn17.70$&$0.507$&$      397.7/   53$&$$&$   \phn\phn17.34$&$0.496$&$      359.9/   52$\\
$13.50$                 &$3.3$&All                 &$45$&$1.00$                 &$      203044$&$      \phn11439$&$   \phn\phn17.75$&$0.550$&$      \phn452.0/   48$&$$&$   \phn\phn17.75$&$0.499$&$      181.3/   47$&$$&$   \phn\phn17.42$&$0.490$&$      165.2/   46$\\
$13.50$                 &$3.3$&All                 &$70$&$1.00$                 &$   \phn39548$&$   \phn\phn2223$&$   \phn\phn17.79$&$0.555$&$      \phn187.1/   32$&$$&$   \phn\phn17.79$&$0.488$&$   \phn92.6/   31$&$$&$   \phn\phn17.29$&$0.472$&$   \phn85.7/   30$\\
\multicolumn{18}{l}{For Varying $ \theta$}\\
$13.50$                 &$3.3$&All                 &$20$&$0.25$                 &$      439754$&$         412101$&$\phn\phn\phn1.07$&$0.258$&$         2266.4/   19$&$$&$\phn\phn\phn1.07$&$0.217$&$      975.6/   18$&$$&$\phn\phn\phn1.05$&$0.212$&$      897.9/   17$\\
$13.50$                 &$3.3$&All                 &$20$&$0.50$                 &$      430439$&$      \phn98048$&$\phn\phn\phn4.39$&$0.406$&$         2069.1/   31$&$$&$\phn\phn\phn4.39$&$0.349$&$      774.9/   30$&$$&$\phn\phn\phn4.29$&$0.338$&$      688.3/   29$\\
$13.50$\tablenotemark{a}&$3.3$&All                 &$20$&$1.00$                 &$      413339$&$      \phn23348$&$   \phn\phn17.70$&$0.565$&$         1126.1/   54$&$$&$   \phn\phn17.70$&$0.507$&$      397.7/   53$&$$&$   \phn\phn17.34$&$0.496$&$      359.9/   52$\\
$13.50$\tablenotemark{a}&$3.3$&All                 &$20$&$2.00$                 &$      384240$&$   \phn\phn5409$&$   \phn\phn71.04$&$0.697$&$      \phn272.8/   93$&$$&$   \phn\phn71.04$&$0.662$&$      138.0/   92$&$$&$   \phn\phn69.81$&$0.655$&$      126.1/   91$\\
$13.50$\tablenotemark{a}&$3.3$&All                 &$20$&$4.00$                 &$      330298$&$   \phn\phn1159$&$      \phn284.99$&$0.793$&$   \phn\phn91.7/   43$&$$&$      \phn284.99$&$0.771$&$   \phn61.7/   42$&$$&$      \phn283.80$&$0.770$&$   \phn61.4/   41$\\
$13.50$                 &$3.3$&All                 &$20$&$6.00$                 &$      280916$&$\phn\phn\phn435$&$      \phn645.78$&$0.835$&$   \phn\phn13.4/   17$&$$&$      \phn645.78$&$0.828$&$   \phn11.7/   16$&$$&$      \phn643.56$&$0.828$&$   \phn11.6/   15$\\
$13.50$                 &$3.3$&All                 &$20$&$8.00$                 &$      231854$&$\phn\phn\phn201$&$         1153.50$&$0.858$&$\phn\phn\phn3.5/\phn7$&$$&$         1153.50$&$0.867$&$\phn\phn2.6/\phn6$&$$&$         1154.00$&$0.867$&$\phn\phn2.6/\phn5$\\
\multicolumn{18}{l}{For Varying Regions of the Sky Using $\theta= 1 \degr$ Cells}\\
$13.50$\tablenotemark{a}&$3.3$&All                 &$20$&$1.00$                 &$      413339$&$      \phn23348$&$   \phn\phn17.70$&$0.565$&$         1126.1/   54$&$$&$   \phn\phn17.70$&$0.507$&$      397.7/   53$&$$&$   \phn\phn17.34$&$0.496$&$      359.9/   52$\\
$13.50$                 &$3.3$&All                 &$20$&$1.00$\tablenotemark{b}&$      413308$&$      \phn23365$&$   \phn\phn17.69$&$0.562$&$      \phn995.5/   55$&$$&$   \phn\phn17.69$&$0.508$&$      352.6/   54$&$$&$   \phn\phn17.33$&$0.498$&$      314.7/   53$\\
$13.50$                 &$3.3$&All\tablenotemark{c}&$20$&$1.00$                 &$      404863$&$      \phn23050$&$   \phn\phn17.56$&$0.554$&$      \phn951.0/   54$&$$&$   \phn\phn17.56$&$0.502$&$      374.4/   53$&$$&$   \phn\phn17.22$&$0.492$&$      341.0/   52$\\
$13.50$                 &$3.3$&N                   &$20$&$1.00$                 &$      217041$&$      \phn12220$&$   \phn\phn17.76$&$0.573$&$      \phn681.0/   49$&$$&$   \phn\phn17.76$&$0.512$&$      221.7/   48$&$$&$   \phn\phn17.39$&$0.502$&$      202.0/   47$\\
$13.50$                 &$3.3$&S                   &$20$&$1.00$                 &$      196298$&$      \phn11128$&$   \phn\phn17.64$&$0.553$&$      \phn501.7/   48$&$$&$   \phn\phn17.64$&$0.500$&$      231.9/   47$&$$&$   \phn\phn17.24$&$0.489$&$      210.5/   46$\\
$13.50$                 &$3.3$&E                   &$20$&$1.00$                 &$      201571$&$      \phn11465$&$   \phn\phn17.58$&$0.561$&$      \phn523.1/   48$&$$&$   \phn\phn17.58$&$0.509$&$      221.0/   47$&$$&$   \phn\phn17.21$&$0.498$&$      202.8/   46$\\
$13.50$                 &$3.3$&W                   &$20$&$1.00$                 &$      210012$&$      \phn11786$&$   \phn\phn17.82$&$0.566$&$      \phn643.1/   49$&$$&$   \phn\phn17.82$&$0.504$&$      203.8/   48$&$$&$   \phn\phn17.43$&$0.493$&$      181.6/   47$\\
$13.50$                 &$3.3$&NW                  &$20$&$1.00$                 &$      111595$&$   \phn\phn6219$&$   \phn\phn17.97$&$0.562$&$      \phn327.6/   42$&$$&$   \phn\phn17.97$&$0.501$&$      103.4/   41$&$$&$   \phn\phn17.65$&$0.494$&$   \phn97.1/   40$\\
$13.50$                 &$3.3$&NW\tablenotemark{c} &$20$&$1.00$                 &$      103119$&$   \phn\phn5921$&$   \phn\phn17.42$&$0.520$&$      \phn327.6/   39$&$$&$   \phn\phn17.42$&$0.483$&$   \phn76.6/   38$&$$&$   \phn\phn17.26$&$0.479$&$   \phn47.7/   37$\\
$13.50$                 &$3.3$&NE                  &$20$&$1.00$                 &$      104720$&$   \phn\phn5958$&$   \phn\phn17.58$&$0.584$&$      \phn370.9/   41$&$$&$   \phn\phn17.58$&$0.523$&$      138.4/   40$&$$&$   \phn\phn17.11$&$0.509$&$      124.3/   39$\\
$13.50$                 &$3.3$&SE                  &$20$&$1.00$                 &$   \phn96851$&$   \phn\phn5507$&$   \phn\phn17.59$&$0.532$&$      \phn198.5/   40$&$$&$   \phn\phn17.59$&$0.493$&$      131.1/   39$&$$&$   \phn\phn17.28$&$0.484$&$      124.4/   38$\\
$13.50$                 &$3.3$&SW                  &$20$&$1.00$                 &$   \phn98417$&$   \phn\phn5567$&$   \phn\phn17.68$&$0.570$&$      \phn362.7/   42$&$$&$   \phn\phn17.68$&$0.505$&$      142.0/   41$&$$&$   \phn\phn17.16$&$0.490$&$      124.0/   40$\\
\multicolumn{18}{l}{For Varying Regions of the Sky Using $\theta= 4 \degr$ Cells}\\
$13.50$\tablenotemark{a}&$3.3$&All                 &$20$&$4.00$                 &$      330298$&$   \phn\phn1159$&$      \phn284.99$&$0.793$&$   \phn\phn91.7/   43$&$$&$      \phn284.99$&$0.771$&$   \phn61.7/   42$&$$&$      \phn283.81$&$0.770$&$   \phn61.4/   41$\\
$13.50$                 &$3.3$&NW                  &$20$&$4.00$                 &$   \phn95024$&$\phn\phn\phn328$&$      \phn289.71$&$0.797$&$   \phn\phn16.2/   12$&$$&$      \phn289.71$&$0.776$&$   \phn11.3/   11$&$$&$      \phn286.57$&$0.773$&$   \phn10.8/   10$\\
$13.50$                 &$3.3$&NE                  &$20$&$4.00$                 &$   \phn83880$&$\phn\phn\phn294$&$      \phn285.31$&$0.817$&$\phn\phn\phn8.1/   11$&$$&$      \phn285.31$&$0.802$&$\phn\phn5.4/   10$&$$&$      \phn282.30$&$0.799$&$\phn\phn5.0/\phn9$\\
$13.50$                 &$3.3$&SE                  &$20$&$4.00$                 &$   \phn71567$&$\phn\phn\phn253$&$      \phn282.87$&$0.756$&$\phn\phn\phn2.7/\phn9$&$$&$      \phn282.87$&$0.747$&$\phn\phn2.4/\phn8$&$$&$      \phn279.20$&$0.742$&$\phn\phn1.7/\phn7$\\
$13.50$                 &$3.3$&SW                  &$20$&$4.00$                 &$   \phn73875$&$\phn\phn\phn264$&$      \phn279.83$&$0.783$&$   \phn\phn14.8/   10$&$$&$      \phn279.83$&$0.771$&$   \phn13.7/\phn9$&$$&$      \phn275.24$&$0.765$&$   \phn12.7/\phn8$
\enddata
\tablecomments{The units of $n_{st}$ are deg$^{-2}$. The units of $\delta_{Gal}$ and $\theta$ are deg.}
\tablenotetext{a}{This row is a parent sample for randomly selected
subsamples in Table~\ref{tbl:GQED_rand}.}
\tablenotetext{b}{The Galactic longitude bins were shifted by $0.5\degr$ for this fit.}
\tablenotetext{c}{Cells containing galaxies within 10 degrees the Shapley Supercluster were excluded for this fit.}
\end{deluxetable}
\clearpage
\end{landscape}
\newpage

\setlength{\tabcolsep}{6pt}
\clearpage
\begin{landscape}
\begin{deluxetable}{cccccccccccccccc}
\tablewidth{0pt}
\tabletypesize{\footnotesize}
\tablecaption{Comparison of Magnitude Selected Subsamples with Randomly Selected Subsamples\label{tbl:GQED_rand}}
\tablehead{
Max&
Poisson&
&
Total&
Total&
\multicolumn{3}{c}{From Variance}&
&
\multicolumn{3}{c}{Fitting $b$}&
&
\multicolumn{3}{c}{Fitting $\overline{N}$ \& $b$}
\\
\cline{6-8} \cline{10-12}  \cline{14-16}\\[-2ex]
\colhead{$K_{s0}$}&
\colhead{Prob.}&
\colhead{$\theta$}&
\colhead{Galaxies}&
\colhead{Cells}&
\colhead{$\overline{N}$}&
\colhead{$b$}&
\colhead{$\chi^{2}$/Dof}&
&
\colhead{$\overline{N}$}&
\colhead{$b$}&
\colhead{$\chi^{2}$/Dof}&
&
\colhead{$\overline{N}$}&
\colhead{$b$}&
\colhead{$\chi^{2}$/Dof}\\
&
}
\startdata
\multicolumn{16}{l}{For Varying  $K_{s0}$ Cutoffs Compared to Random Selection from $K_{s0}=13.50 $ Cutoff  Using $\theta= 1 \degr$ Cells}\\
$11.00$                 &$1.0000$&$1.00$                 &$   \phn10940$&$      \phn23348$&$\phn\phn\phn0.47$&$0.264$&$   \phn\phn84.5/\phn8$&$$&$\phn\phn\phn0.47$&$0.223$&$   \phn37.6/\phn7$&$$&$\phn\phn\phn0.46$&$0.218$&$   \phn34.8/\phn6$\\
$13.50$                 &$0.0265$&$1.00$                 &$   \phn10917$&$      \phn23348$&$\phn\phn\phn0.47$&$0.049$&$   \phn\phn12.1/\phn5$&$$&$\phn\phn\phn0.47$&$0.044$&$   \phn11.2/\phn4$&$$&$\phn\phn\phn0.47$&$0.043$&$   \phn11.1/\phn3$\\
$11.50$                 &$1.0000$&$1.00$                 &$   \phn22252$&$      \phn23348$&$\phn\phn\phn0.95$&$0.329$&$      \phn250.8/   12$&$$&$\phn\phn\phn0.95$&$0.266$&$   \phn87.3/   11$&$$&$\phn\phn\phn0.92$&$0.257$&$   \phn76.7/   10$\\
$13.50$                 &$0.0538$&$1.00$                 &$   \phn22343$&$      \phn23348$&$\phn\phn\phn0.96$&$0.091$&$   \phn\phn13.3/\phn8$&$$&$\phn\phn\phn0.96$&$0.085$&$   \phn11.4/\phn7$&$$&$\phn\phn\phn0.95$&$0.084$&$   \phn11.3/\phn6$\\
$12.00$                 &$1.0000$&$1.00$                 &$   \phn45385$&$      \phn23348$&$\phn\phn\phn1.94$&$0.388$&$      \phn479.2/   17$&$$&$\phn\phn\phn1.94$&$0.317$&$      166.8/   16$&$$&$\phn\phn\phn1.88$&$0.305$&$      144.5/   15$\\
$13.50$                 &$0.1100$&$1.00$                 &$   \phn45530$&$      \phn23348$&$\phn\phn\phn1.95$&$0.188$&$   \phn\phn87.5/   12$&$$&$\phn\phn\phn1.95$&$0.158$&$   \phn45.2/   11$&$$&$\phn\phn\phn1.93$&$0.157$&$   \phn43.2/   10$\\
$12.50$                 &$1.0000$&$1.00$                 &$   \phn94467$&$      \phn23348$&$\phn\phn\phn4.05$&$0.453$&$      \phn731.8/   24$&$$&$\phn\phn\phn4.05$&$0.382$&$      243.3/   23$&$$&$\phn\phn\phn3.92$&$0.368$&$      209.6/   22$\\
$13.50$                 &$0.2290$&$1.00$                 &$   \phn94631$&$      \phn23348$&$\phn\phn\phn4.05$&$0.292$&$      \phn278.1/   19$&$$&$\phn\phn\phn4.05$&$0.246$&$      142.6/   18$&$$&$\phn\phn\phn4.00$&$0.241$&$      132.2/   17$\\
$13.00$                 &$1.0000$&$1.00$                 &$      197738$&$      \phn23348$&$\phn\phn\phn8.47$&$0.514$&$         1108.0/   37$&$$&$\phn\phn\phn8.47$&$0.446$&$      364.2/   36$&$$&$\phn\phn\phn8.24$&$0.433$&$      326.3/   35$\\
$13.50$                 &$0.4780$&$1.00$                 &$      197645$&$      \phn23348$&$\phn\phn\phn8.47$&$0.425$&$      \phn604.3/   33$&$$&$\phn\phn\phn8.47$&$0.371$&$      262.7/   32$&$$&$\phn\phn\phn8.31$&$0.362$&$      239.9/   31$\\
\multicolumn{16}{l}{For Varying  $K_{s0}$ Cutoffs Compared to Random Selection from $K_{s0}=13.50 $ Cutoff  Using $\theta= 2 \degr $ Cells}\\
$11.00$                 &$1.0000$&$2.00$                 &$   \phn10111$&$   \phn\phn5409$&$\phn\phn\phn1.87$&$0.390$&$   \phn\phn49.3/   12$&$$&$\phn\phn\phn1.87$&$0.352$&$   \phn29.6/   11$&$$&$\phn\phn\phn1.82$&$0.344$&$   \phn27.3/   10$\\
$13.50$                 &$0.0263$&$2.00$                 &$   \phn10279$&$   \phn\phn5409$&$\phn\phn\phn1.90$&$0.112$&$   \phn\phn10.3/\phn8$&$$&$\phn\phn\phn1.90$&$0.102$&$\phn\phn9.2/\phn7$&$$&$\phn\phn\phn1.90$&$0.102$&$\phn\phn9.1/\phn6$\\
$11.50$                 &$1.0000$&$2.00$                 &$   \phn20585$&$   \phn\phn5409$&$\phn\phn\phn3.81$&$0.465$&$   \phn\phn85.9/   18$&$$&$\phn\phn\phn3.81$&$0.418$&$   \phn38.2/   17$&$$&$\phn\phn\phn3.71$&$0.409$&$   \phn34.5/   16$\\
$13.50$                 &$0.0536$&$2.00$                 &$   \phn20859$&$   \phn\phn5409$&$\phn\phn\phn3.86$&$0.179$&$   \phn\phn25.3/   13$&$$&$\phn\phn\phn3.86$&$0.154$&$   \phn16.9/   12$&$$&$\phn\phn\phn3.84$&$0.152$&$   \phn16.5/   11$\\
$12.00$                 &$1.0000$&$2.00$                 &$   \phn42092$&$   \phn\phn5409$&$\phn\phn\phn7.78$&$0.537$&$      \phn205.3/   27$&$$&$\phn\phn\phn7.78$&$0.481$&$   \phn95.8/   26$&$$&$\phn\phn\phn7.53$&$0.467$&$   \phn85.7/   25$\\
$13.50$                 &$0.1100$&$2.00$                 &$   \phn42268$&$   \phn\phn5409$&$\phn\phn\phn7.81$&$0.309$&$   \phn\phn34.7/   20$&$$&$\phn\phn\phn7.81$&$0.281$&$   \phn20.9/   19$&$$&$\phn\phn\phn7.76$&$0.279$&$   \phn20.0/   18$\\
$12.50$                 &$1.0000$&$2.00$                 &$   \phn87685$&$   \phn\phn5409$&$   \phn\phn16.21$&$0.603$&$      \phn297.3/   41$&$$&$   \phn\phn16.21$&$0.547$&$      127.1/   40$&$$&$   \phn\phn15.69$&$0.532$&$      111.0/   39$\\
$13.50$                 &$0.2280$&$2.00$                 &$   \phn87784$&$   \phn\phn5409$&$   \phn\phn16.23$&$0.450$&$      \phn162.2/   34$&$$&$   \phn\phn15.23$&$0.401$&$   \phn84.5/   33$&$$&$   \phn\phn16.02$&$0.394$&$   \phn79.7/   32$\\
$13.00$                 &$1.0000$&$2.00$                 &$      183757$&$   \phn\phn5409$&$   \phn\phn33.97$&$0.656$&$      \phn342.5/   61$&$$&$   \phn\phn33.97$&$0.610$&$      164.9/   60$&$$&$   \phn\phn33.15$&$0.599$&$      150.2/   59$\\
$13.50$                 &$0.4780$&$2.00$                 &$      183759$&$   \phn\phn5409$&$   \phn\phn33.97$&$0.579$&$      \phn203.4/   56$&$$&$   \phn\phn33.97$&$0.539$&$      114.1/   55$&$$&$   \phn\phn33.43$&$0.532$&$      104.5/   54$\\
\multicolumn{16}{l}{For Varying  $K_{s0}$ Cutoffs Compared to Random Selection from $K_{s0}=13.50 $ Cutoff  Using $\theta= 4 \degr $ Cells}\\
$11.00$                 &$1.0000$&$4.00$                 &$\phn\phn8509$&$   \phn\phn1159$&$\phn\phn\phn7.34$&$0.546$&$   \phn\phn37.3/   17$&$$&$\phn\phn\phn7.34$&$0.504$&$   \phn22.7/   16$&$$&$\phn\phn\phn7.19$&$0.496$&$   \phn21.9/   15$\\
$13.50$                 &$0.0258$&$4.00$                 &$\phn\phn8517$&$   \phn\phn1159$&$\phn\phn\phn7.35$&$0.217$&$   \phn\phn12.4/   14$&$$&$\phn\phn\phn7.35$&$0.203$&$   \phn11.8/   13$&$$&$\phn\phn\phn7.32$&$0.202$&$   \phn11.7/   12$\\
$11.50$                 &$1.0000$&$4.00$                 &$   \phn17430$&$   \phn\phn1159$&$   \phn\phn15.04$&$0.610$&$   \phn\phn55.1/   24$&$$&$   \phn\phn15.04$&$0.575$&$   \phn40.3/   23$&$$&$   \phn\phn14.63$&$0.564$&$   \phn38.2/   22$\\
$13.50$                 &$0.0528$&$4.00$                 &$   \phn17385$&$   \phn\phn1159$&$   \phn\phn15.00$&$0.325$&$   \phn\phn23.2/   20$&$$&$   \phn\phn15.00$&$0.304$&$   \phn21.2/   19$&$$&$   \phn\phn14.93$&$0.301$&$   \phn21.0/   18$\\
$12.00$                 &$1.0000$&$4.00$                 &$   \phn35665$&$   \phn\phn1159$&$   \phn\phn30.77$&$0.674$&$   \phn\phn96.1/   35$&$$&$   \phn\phn30.77$&$0.633$&$   \phn64.0/   34$&$$&$   \phn\phn29.95$&$0.622$&$   \phn60.9/   33$\\
$13.50$                 &$0.1080$&$4.00$                 &$   \phn35619$&$   \phn\phn1159$&$   \phn\phn30.73$&$0.449$&$   \phn\phn53.2/   29$&$$&$   \phn\phn30.73$&$0.424$&$   \phn49.1/   28$&$$&$   \phn\phn30.47$&$0.420$&$   \phn48.3/   27$\\
$12.50$                 &$1.0000$&$4.00$                 &$   \phn74812$&$   \phn\phn1159$&$   \phn\phn64.55$&$0.725$&$   \phn\phn78.9/   35$&$$&$   \phn\phn64.55$&$0.689$&$   \phn40.9/   34$&$$&$   \phn\phn63.61$&$0.683$&$   \phn39.6/   33$\\
$13.50$                 &$0.2260$&$4.00$                 &$   \phn74913$&$   \phn\phn1159$&$   \phn\phn64.64$&$0.591$&$   \phn\phn57.8/   33$&$$&$   \phn\phn64.64$&$0.557$&$   \phn41.7/   32$&$$&$   \phn\phn64.45$&$0.555$&$   \phn41.6/   31$\\
$13.00$                 &$1.0000$&$4.00$                 &$      157519$&$   \phn\phn1159$&$      \phn135.91$&$0.764$&$   \phn\phn77.1/   39$&$$&$      \phn135.91$&$0.735$&$   \phn40.4/   38$&$$&$      \phn134.46$&$0.731$&$   \phn39.4/   37$\\
$13.50$                 &$0.4770$&$4.00$                 &$      157197$&$   \phn\phn1159$&$      \phn135.63$&$0.708$&$   \phn\phn66.5/   39$&$$&$      \phn135.63$&$0.680$&$   \phn43.8/   38$&$$&$      \phn134.58$&$0.677$&$   \phn42.9/   37$\\
\enddata
\tablecomments{All fits were to All-sky data where $n_{st}<3.3$ deg$^{-2}$ and $\vert \delta_{Gal} \vert > 20$ deg.
The units of $\theta$ are deg.}
\end{deluxetable}
\clearpage
\end{landscape}

\end{document}